\documentclass[aps,prl,floatfix,superscriptaddress,twocolumn,showpacs]{revtex4}
\usepackage{times}
\usepackage{dcolumn}                 
\newcolumntype{w}[1]{D{.}{.}{#1}}
\begin{document}

\title{Improved theory of helium fine structure}

\author{Krzysztof Pachucki}

\affiliation{
Institute of Theoretical Physics, Warsaw University,
Ho\.{z}a 69, 00-681 Warsaw, Poland} 

\begin{abstract}
Improved theoretical predictions for the fine-structure splitting
of $2^3P_J$ levels in helium are obtained by the calculation of 
contributions of order $\alpha^5\,$Ry.
New results for transition frequencies, 
$\nu_{01} = 29\,616\,943.01(17)$ kHz, and $\nu_{12} =  2\,291\,161.13(30)$ kHz,
disagree significantly with the experimental values, indicating an outstanding 
problem in bound state QED. 
\end{abstract}


\maketitle

The fine-structure splitting of the helium $2^3 P_J$ states is an 
intrinsically relativistic effect, and arises from
the interaction of spins and orbital angular momentum.
The value of this splitting has been measured 
with increasing precision over the last years
\cite{gab,Hessels1,Hessels2, Inguscio1, Inguscio2, shiner}.
Since the splitting is proportional to $\alpha^2$ Ry, these accurate
measurements make helium a candidate for determining the fine structure
constant $\alpha$, provided that the higher order in $\alpha$ corrections can
be sufficiently well understood. 
The most accurate determination of $\alpha$ at present comes from the $g-2$ of
the electron. This determination depends sensitively 
on complicated multi-loop calculations performed by Kinoshita and by Remiddi 
and coworkers \cite{mohr}, and therefore requires independent confirmation. 
In response to  significant experimental
effort \cite{gab,Hessels1,Hessels2, Inguscio1,Inguscio2, shiner}, we present here the calculation of 
the $\alpha^5$ Ry contribution to helium fine structure, so that these experiments can
be used to provide an independent determination of $\alpha$.

Several recent advances in bound state Quantum Electrodynamics (QED) have
made the calculation of higher order corrections to helium fine structure possible.
Specifically, Yelkhovsky in Ref. \cite{yel} has shown how to use dimensional 
regularization in the calculation of helium energy
levels, and together with Korobov has obtained in \cite{kor} numerical values for
the $\alpha^4$ Ry contributions to the ground state.
Next, in Ref. \cite{nrqed} a Foldy-Wouthuysen transformed QED Langrangian was
used to derive all effective $\alpha^4$ Ry operators
for arbitrary states of few electron atoms.
More recently, together with Jentschura and Czarnecki, 
we have obtained in Ref. \cite{lamb} general formulae for $\alpha^5$ Ry 
correction to hydrogenic energy levels, including the fine structure.
The calculational approach of these works \cite{kor, lamb} and the present
paper is based on dimensionally 
regularized QED. The parameter
$\epsilon$, related to the space dimension $d=3-2\,\epsilon$,
plays the role of both infrared and ultraviolet regulator,
as some $\alpha^5$ Ry terms are divergent in $d=3$ space. 
This artificial parameter $\epsilon$ is used to derive various terms, 
and we will explicitly demonstrate its cancellation in their sum.
Natural relativistic units will be used with 
$\hbar = c = \epsilon_0= m =1$, so that $e^2 = 4\,\pi\,\alpha$.

The fine structure in order $m\,\alpha^7$ ($\alpha^5\,$Ry) can be written as 
\cite{Pach2}
\begin{eqnarray}
\hspace*{-1ex}E^{(7)}&=& \langle H^{(7)}\rangle+
2\,\left\langle H^{(4)}\,\frac{1}{(E_0-H_0)'}\,H^{(5)}\right\rangle+E_{L}
\label{02}
\end{eqnarray} 
where $E_{L}$ is the Bethe logarithmic correction of Eq. (\ref{bethe}), 
and $H^{(i)}$ is an effective Hamiltonian
of order $m\,\alpha^i$. We will concentrate in this work on a complete 
derivation of $H^{(7)}$, as the other terms contributing to order
$m\,\alpha^7$ ($\alpha^5\,$Ry), $E_L$ and the second 
order term called $E_S$, have already been obtained in \cite{Pach2}. Important
terms of order $m\,\alpha^7\,\ln\alpha$ first calculated in Ref. \cite{log}
are confirmed in the present calculation.
$H^{(7)}$ consists of  exchange terms and radiative
corrections, where a photon is emitted and absorbed by the same particle.
We consider first the exchange terms. Their derivation in
general is quite complicated. We note that only two-photon
exchange diagrams contribute and there are no three-body terms, 
which is a result of an internal cancellation. 
A feature of the calculation that leads to considerable simplification is the fact that
the order being calculated in nonanalytic in $\alpha^2$.
For example, $H^{(5)}$ includes only two terms
\begin{equation}
H^{(5)} = - \frac{7}{6\,\pi}\,\frac{\alpha^2}{r^3}
+ \frac{38\,Z\,\alpha^2}{45}\,\left[\delta^3(r_1)+\delta^3(r_2)\right]\,, 
\label{18}
\end{equation}
and they can be derived from the two-photon exchange scattering amplitude.
Similar results hold for the spin dependent $m\,\alpha^7$ terms.
If $H^{(7)}$ represents an effective Hamiltonian, it has to give the same
scattering amplitude as in full QED. Therefore, 
we obtain the exchange contribution $\delta H$ from the spin dependent 
part of the two-photon scattering amplitude, which is
\begin{eqnarray}
\delta_1 H &=& \frac{i\,e^4}{(2\,\pi)^D}\,\int d^D k\,
\frac{1}{(k+q/2)^2}\,\frac{1}{(k-q/2)^2}\nonumber \nonumber \\ &&
\biggl[\bar u(p'_1)\,\gamma^\mu\,\frac{1}
{\not\!k+(\not\!p_1 + \not\!p'_1)/2-1}\,\gamma^\nu\,u(p_1) \nonumber \\ &&
+\bar u(p'_1)\,\gamma^\nu\,\frac{1}
{-\not\!k+(\not\!p_1 + \not\!p'_1)/2-1}\,\gamma^\mu\,u(p_1)\biggr]
\nonumber \\ &&
\times \bar u(p'_2)\,\gamma^\nu\,\frac{1}
{\not\!k+(\not\!p_2 + \not\!p'_2)/2-1}\,\gamma^\mu\,u(p_2)
\label{03}
\end{eqnarray}
where $q = p'_1 - p_1 = p_2-p'_2$. 
If one expands this amplitude in small external momenta one obtains
\begin{eqnarray}
\delta_1 H &=& \alpha^2\biggl\{ 
\sigma_1(j,q)\,\sigma_2(j,q)\,\left[-\frac{19}{18} 
+\frac{1}{3\,\epsilon}\,+ \frac{1}{2}\,\ln (q)\right]
 \nonumber\\ &&  
+i\,[\sigma_1({p'_1},{p_1})+\sigma_2({p'_2},{p_2})]\,
\left[ \frac{5}{12}- \frac{1}{3\,\epsilon} + \frac{1}{6}\,\ln (q)\right]
\nonumber \\ && 
+i\,[\sigma_1({p'_2},{p_2})+\sigma_2({p'_1},{p_1})]\!
\left[\frac{11}{12} - \frac{2}{3\,\epsilon} + \frac{4}{3}\,\ln (q)\right]
 \nonumber\\ && 
  +\frac{1}{8}\,\sigma_1(j,{p_1} + {p'_1})\,
      \sigma_2(j,{p_2} + {p'_2})
 \nonumber\\ &&
  -\frac{1}{8}\,\sigma_1(j,{p_2} + {p'_2})\,
        \sigma_2(j,{p_1} + {p'_1})
 \nonumber\\ && 
+\frac{17}{72}\,\sigma_1(j,{p_1} - {p_2} + {p'_1} - {p'_2})
\nonumber\\ && 
      \times\sigma_2(j,{p_1} - {p_2} + {p'_1} - {p'_2}) \biggr\}
\label{05}
\end{eqnarray} 
where $\sigma^{ij} = -i/2\,[\sigma^i,\sigma^j]$ and $\sigma(j,q) =\sigma^{ji}\,q^i$.
The $1/\epsilon$ divergences cancel out with the low energy part 
where photon momenta are of the order of the binding energy.
This low energy contribution gives the Bethe logarithm,
described later in Eq. (\ref{bethe}), and the correction 
\begin{eqnarray}
\delta E_L &=& e^2\,\int_\Lambda^{\infty}
\,\frac{d^dk}{(2\,\pi)^d\,2\,k}\,
\left(\delta^{ij}-\frac{k^i\,k^j}{k^2}\right)
\nonumber \\ &&\times
\delta\left\langle\phi\left|p_1^i\,\frac{1}{E-H-k}\,p_2^j
\right|\phi\right\rangle + (1 \leftrightarrow 2),
\label{06}
\end{eqnarray}
which is the transition term from dimensional regularization to
the direct $\Lambda = m\,(Z\,\alpha)^2\,\lambda$ cut-off in the photon momenta.
Here $\delta$ denotes the first order correction to $\phi$, $H$ and $E$ due to 
the spin dependent part of the Breit-Pauli Hamiltonian $H^{(4)}$.
The resulting correction is a sum of two terms. The first one
contributes to $\langle H^{(4)} /(E_0-H_0)'\,H^{(5)}\rangle$
in Eq. (\ref{02}), and the second term is the effective Hamiltonian
\begin{eqnarray}
\delta_2 H &=& \alpha^2\left[\frac{5}{9}+\frac{1}{3\,\epsilon}+
\frac{2}{3}\,\ln[(Z\,\alpha)^{-2}]\right]\,
[i\,\sigma_1({p'_1},{p_1})
\nonumber \\ &&
+i\,\sigma_2({p'_2},{p_2})+2\,i\,\sigma_1({p'_2},{p_2})
+2\,i\,\sigma_2({p'_1},{p_1})
\nonumber \\ &&
-\sigma_1(j,q)\,\sigma_2(j,q)],
\label{07}
\end{eqnarray}
where we omitted a $\ln2\lambda$ term.
Together with Eq. (\ref{05}) this gives the complete contribution due to
exchange terms. When calculating expectation values on $^3P_J$ 
states further simplifications can be performed. Namely,
the expectation value of a Dirac delta function with both momenta
on the right or on the left hand side vanishes. Moreover,
the nonrelativistic wave function is a product of a symmetric spin 
and an antisymmetric spatial function. This means that the expectation value
of $\sigma_1$ is equal to that of $\sigma_2$. As a result the 
total exchange contribution $H_E = \delta_1 H + \delta_2 H$ is
\begin{eqnarray}
H_E &=&  \alpha^2\,\biggl[6+4\ln[(Z\,\alpha)^{-2}]
+3\,\ln q\biggr]i\,\sigma_1({p'_1},{p_1})
\nonumber \\ && 
+\alpha^2\,\biggl[-\frac{23}{9}
-\frac{2}{3}\,\ln[(Z\,\alpha)^{-2}]
+\frac{1}{2}\,\ln q\biggr]
\nonumber \\ &&
{\scriptstyle \times}\sigma_1(j,q)\,\sigma_2(j,q).
\label{08}
\end{eqnarray}

The treatment of the radiative correction is different. We argue
that radiative corrections can be incorporated by the use of
electromagnetic formfactors and a Uehling correction to the Coulomb potential
\begin{eqnarray}
F_1(-\vec q^{\,2}) &=&
1+\frac{\alpha}{\pi}\left(\frac{1}{8}+\frac{1}{6\,\epsilon}\right)\,\vec q^{\,2}\nonumber \\
F_2(-\vec q^{\,2}) &=& \frac{\alpha}{\pi}\left(\frac{1}{2}-\frac{1}{12}\,\vec q^{\,2}\right)\nonumber \\
F_V(-\vec q^{\,2}) &=&  \frac{\alpha}{\pi}\,\frac{1}{15}\,\vec q^{\,2}
\label{09}
\end{eqnarray}
The possible additional corrections are quadratic
in electromagnetic fields: see Ref. \cite{lamb}. However, 
terms formed out of $\vec E, \vec B, \vec p,\vec\sigma$
can contribute only at higher order and thus can be neglected.
Corrections due to the slope of formfactors and the vacuum polarization
are obtained analogously to  the Breit-Pauli Hamiltonian $H^{(4)}$,
by modifying electromagnetic vertices and the photon propagator.
The result is 
\begin{eqnarray}
\delta_3 H &=& \pi\,Z\,\alpha(F'_1+2\,F'_2+F_V')
i\,[\sigma_1(p_1'',p_1) + \sigma_2(p_2'',p_2)]
\nonumber \\ && -
\pi\,\alpha(2\,F'_1+2\,F'_2+F_V')
i\,[\sigma_1(p_1',p_1) + \sigma_2(p_2',p_2)]
\nonumber \\ &&
-2\,\pi\,\alpha(2\,F'_1+F'_2+F_V')
i\,[\sigma_1(p_2',p_2) + \sigma_2(p_1',p_1)]
\nonumber \\ &&
+\pi\,\alpha(2\,F'_1+2\,F'_2+F_V')\,
\sigma_1(j,q)\,\sigma_1(j,q)\,,
\label{10}
\end{eqnarray}
where by $p''$ we denote momentum scattered off 
the Coulomb potential of a nucleus, and $F' = F'(0)$.
There is also a low-energy contribution which is calculated 
in a way similar to this in Eq. (\ref{06}), namely
\begin{eqnarray}
\delta E_L &=& e^2\,\int_\Lambda^{\infty}
\,\frac{d^dk}{(2\,\pi)^d\,2\,k}\,
\left(\delta^{ij}-\frac{k^i\,k^j}{k^2}\right)
\nonumber \\ &&\times
\delta\left\langle\phi\left|p_1^i\,\frac{1}{E-H-k}\,p_1^j
\right|\phi\right\rangle + (1 \rightarrow 2) 
\label{11}
\end{eqnarray}
The resulting effective Hamiltonian is
\begin{eqnarray}
\delta_4 H &=& \alpha^2\,\left[\frac{5}{9}+\frac{1}{3\,\epsilon}
+\frac{2}{3}\,\ln[(Z\,\alpha)^{-2}]\right]
\nonumber \\ &&
\times \biggl[\frac{i\,Z}{2}\,\sigma_1(p''_1,p_1)+
\frac{i\,Z}{2}\,\sigma_2(p''_2,p_2)
\nonumber \\ &&
-i\,\sigma_1(p'_1,p_1) -i\,\sigma_2(p'_2,p_2) 
-2\,i\,\sigma_2(p'_1,p_1) 
\nonumber \\ &&
-2\,i\,\sigma_1(p'_2,p_2)
+\sigma_1(j,q)\,\sigma_2(j,q)\biggr]
\label{12}
\end{eqnarray}  
The complete radiative correction is a sum of Eqs. (\ref{10}) and (\ref{12}),
namely $H_R = \delta_3 H + \delta_4 H$. Using symmetry $1\leftrightarrow 2$ 
it takes the form
\begin{eqnarray}
H_R &=& Z\,\alpha^2\left[\frac{91}{180}+\frac{2}{3}\,
\ln[(Z\,\alpha)^{-2}]\right]i\,\sigma_1(p''_1,p_1) 
\nonumber \\ &&
+\alpha^2\left[\frac{73}{180}+\frac{2}{3}\,\ln[(Z\,\alpha)^{-2}]\right]\,
\sigma_1(j,q)\,\sigma_2(j,q)
\nonumber \\ &&
-\alpha^2\left[\frac{21}{10}+4\,\ln[(Z\,\alpha)^{-2}]\right]\,
i\,\sigma_1(p'_1,p_1)
\label{13}
\end{eqnarray}
It is convenient to consider a sum of Eqs. (\ref{08}) and (\ref{13}),
as several logarithmic terms cancel out
\begin{equation}
H_Q = H_E+H_R = \sum_{i=1}^5 Q_i
\end{equation}
The logarithmic terms agree with Refs. \cite{log,Pach1}, while
nonlogarithmic terms $Q_i$ are presented in Table I.
\begin{table}[!hbt]
\caption{Operators due to exchange diagrams, slope of formfactors and the vacuum
  polarization, in atomic units with a prefactor $m\,\alpha^7/\pi$.
  The singular $\int dr/r$ integral is defined with an implicit lower
  cut-off $\lambda$ and the term $\ln\lambda+\gamma$ is subtracted out. }
\label{table1}
\(
\begin{array}{lrr}
\mbox{\rm Operator} & \nu_{01}[{\rm kHz}]  & \nu_{12}[{\rm kHz}] 
\\ \hline
Q_1 = \frac{91\,\pi}{180}\,Z\,i\,\vec p_1\times\delta^3(r_1)\,\vec p_1\cdot\vec\sigma_1
&2.854      &5.709   \\[1ex]
Q_2 = -\frac{83\,\pi}{60}\,\vec\sigma_1\cdot\vec\nabla\,\vec\sigma_2\cdot\vec\nabla\delta^3(r)
&10.886     &-4.355      \\[1ex]
Q_3 = -\frac{15}{8}\,\frac{1}{r^7}\,\vec r\cdot\vec\sigma_1\,\vec r\cdot\vec\sigma_2
&4.132        &-1.653       \\[1ex]
Q_4 = \frac{69\,\pi}{10}\,i\vec p_1\times\delta^3(r)\vec p_1\cdot\vec\sigma_1
&5.186        &10.372    \\[1ex]
Q_5 = -\frac{3\,i}{4}\,\vec p_1\times\frac{1}{r^3}\,\vec p_1\cdot\vec\sigma_1
&-1.328       &-2.656     \\[1ex] \hline
E_Q = \sum_{i=1,5} Q_i &21.731    & 7.418
\end{array}
\)
\end{table}

The remaining contribution is the anomalous magnetic moment correction to
the spin dependent operators. We derive it
with the help of a nonrelativistic QED Hamiltonian obtained
by a Foldy-Wouthuysen transformation of the Dirac Hamiltonian including
the magnetic moment anomaly $\kappa$ \cite{lamb}
\begin{eqnarray}
H_{FW} &=& \frac{\vec \pi^2}{2}+e\,A^0 -\frac{e}{2}\,(1+\kappa)\,
\vec\sigma\cdot\vec B -\frac{\vec \pi^4}{8}
\nonumber \\ &&
-\frac{e}{8}\,(1+2\,\kappa)\,[\vec\nabla\cdot\vec E+
\vec\sigma\cdot(\vec E\times\vec\pi-\vec\pi\times\vec E)]
\nonumber \\ &&
+\frac{e}{8}\bigl(\{\vec\sigma\cdot\vec B,\vec\pi^2\}
+\kappa\,\{\vec\pi\cdot\vec B, \vec\pi\cdot\vec\sigma\}\bigr)
\nonumber \\ &&
-\frac{(3+4\,\kappa)}{64}\,\{\vec p^{\,2},e\,\vec E\times\vec
p\cdot\vec\sigma\}
\label{15}
\end{eqnarray}
All the $m\,\alpha^6$ operators obtained by Douglas and Kroll (DK) in \cite{DK}
can also be obtained  from this Hamiltonian in Eq. (\ref{15}), see Ref.\cite{nrqed}.
The anomalous magnetic moment operators are derived in a very similar way.
They differ (see Table II) only by multiplicative factors
from the DK operators. There is a one to one correspondence with Table I 
of Ref. \cite{daley} with 3 exceptions. The operator $H_8$ from  Table II
canceled out in DK calculation. The other two exceptions are related to the different 
spin structure of the next to last term in Eq. (\ref{15}), which leads to operators $H_{16}$
and $H_{17}$ in our Table II.
\begin{table}[!hbt]
\caption{ Operators due to magnetic moment anomaly in atomic units
          with the prefactor $m\,\alpha^7/\pi$}
\label{table2}
\(
\begin{array}{lrr}
\mbox{\rm Operator} & \nu_{01}[{\rm kHz}]  & \nu_{12}[{\rm kHz}] \\
\hline \\[-2ex]
H_1 = -\frac{Z}{4}\,p_1^2\,\frac{\vec r_1}{r_1^3}\times\vec p_1\cdot\vec\sigma_1
&3.239        &6.478        \\[1ex]
H_2 = -\frac{3\,Z}{4}\,\frac{\vec r_1}{r_1^3}\times\frac{\vec r}{r^3}
\cdot\vec\sigma_1\,(\vec r\cdot\vec p_2)
&0.267      & 0.534        \\[1ex]
H_3 = \frac{3\,Z}{4}\,\frac{\vec r}{r^3}\cdot\vec\sigma_1\,
\frac{\vec r_1}{r_1^3}\cdot\vec\sigma_2
&0.332        &-0.133         \\[1ex]
H_4 = \frac{1}{2\,r^4}\,\vec r\times\vec p_2\cdot\vec\sigma_1
&0.749        &1.498       \\[1ex]
H_5 = -\frac{3}{4\,r^6}\,\vec r\cdot\vec\sigma_1\,\vec r\cdot\vec\sigma_2
&2.638       &-1.055        \\[1ex]
H_6 = \frac{1}{4}\,p_1^2\,\frac{\vec r}{r^3}\times\vec p_1\cdot\vec\sigma_1
&-0.807       &-1.614         \\[1ex]
H_7 = -\frac{1}{4}\,p_1^2\,\frac{\vec r}{r^3}\times\vec p_2\cdot\vec\sigma_1
&-1.237         &-2.474       \\[1ex]
H_8 = -\frac{Z}{4\,r}\,\frac{\vec r_1}{r_1^3}\times\vec p_2\cdot\vec\sigma_1
&-0.460        &-0.920         \\[1ex]
H_9 = -\frac{i}{2}\,p_1^2\,\frac{1}{r^3}\,\vec r\cdot\vec p_2\,\vec r\times\vec p_1\cdot\vec\sigma_1 
&0.093        &0.187       \\[1ex]
H_{10} = \frac{3\,i}{4\,r^5}\,\vec r\times
       (\vec r\cdot \vec p_2)\,\vec p_1\cdot \vec\sigma_1 
&-0.376       &-0.752        \\[1ex]
H_{11} = -\frac{3}{8\,r^5}\,\vec r\times
       (\vec r\times \vec p_1\cdot\vec\sigma_1)\,\vec p_2\cdot \vec\sigma_2 
&-0.193         &0.077          \\[1ex]
H_{12} = -\frac{1}{8\,r^3}\,\vec p_1\cdot\vec\sigma_2\,\vec p_2\cdot\vec\sigma_1
&-0.447         &0.179       \\[1ex]
H_{13} = \frac{21}{16}\,p_1^2\,\frac{1}{r^5}\,\vec r\cdot\vec\sigma_1\,\vec r\cdot\vec\sigma_2 
&-14.908         &5.963        \\[1ex]
H_{14} = -\frac{3\,i}{8}\,p_1^2\,\frac{\vec r}{r^3}\cdot\vec\sigma_1\,\vec p_1\cdot\vec\sigma_2
&4.411        &-1.764         \\[1ex]
H_{15} = \frac{i}{8}\,p_1^2\,\frac{1}{r^3}\,\bigl(\vec r\cdot\vec\sigma_2\,\vec p_2\cdot\vec\sigma_1
         +(\vec r\cdot\vec \sigma_1) \\[1ex] \hspace*{5ex}
         {\scriptscriptstyle \times}(\vec p_2\cdot\vec\sigma_2) 
         -\frac{3}{r^2}\,\vec r\cdot\vec\sigma_1\,\vec
          r\cdot\vec\sigma_2\,\vec r\cdot\vec p_2\bigr)
&4.618         &-1.847         \\[1ex]
H_{16} = -\frac{1}{4}\,\vec p_1\cdot\vec \sigma_1\,\vec p_1\times\frac{\vec r}{r^3}\cdot\vec p_2
&-0.483         & -0.967        \\[1ex]
H_{17} = \frac{1}{8}\,\vec p_1\cdot\vec\sigma_1\,\bigl(-\vec
p_1\cdot\vec\sigma_2\,\frac{1}{r^3}  \\[1ex]\hspace*{6ex}
+3\vec p_1\cdot\vec r\,\frac{\vec r}{r^5}\cdot\vec\sigma_2\bigr)
&-1.643    & 0.657        
\\[1ex] \hline 
E_H = \sum_{i=1,17} \langle H_i\rangle &-4.208    & 4.047 
\end{array}
\)
\end{table}

Apart from the $H_i$ and $Q_i$ operators, second order contributions and
low energy Bethe-logarithmic type corrections contribute to the fine
structure. These contributions have already been
considered in our former work \cite{Pach2}. The second order contribution
$E_S$, beyond the anomalous magnetic moment terms is the second term of Eq. (\ref{02}),
and the low energy contribution $E_L$ is
\begin{eqnarray}
E_{L} &=&-\frac{2\,\alpha}{3\,\pi}\,
\delta\,\biggl\langle \phi\biggl|\,(\vec p_1+\vec p_2)\,(H-E)
\ln\left[\frac{2(H-E)}{(Z\,\alpha)^2}\right]\nonumber \\ &&
(\vec p_1+\vec p_2)\,\biggr|\phi\biggr\rangle
+\frac{i\,Z^2\,\alpha^3}{3\,\pi}
\,\biggl\langle\phi\biggl|
\left(\frac{\vec r_1}{r_1^3}+\frac{\vec r_2}{r_2^3}\right)\label{bethe}
\\ &&
\times\frac{(\vec \sigma_1+\vec\sigma_2)}{2}
\ln\left[\frac{2(H-E)}{(Z\,\alpha)^2}\right]\cdot
\left(\frac{\vec r_1}{r_1^3}+\frac{\vec r_2}{r_2^3}\right)
\biggr|\phi\biggr\rangle, \nonumber 
\end{eqnarray}
where $\delta\langle\ldots\rangle$ denotes the correction to the matrix element
$\langle\ldots\rangle$ due to $H^{(4)}$.
Numerical results for all these contributions is presented in Table III.
\begin{table}[!hbt]
\caption{Summary of contributions to helium fine structure, $E^{(4)}$ and
  $E^{(6)}$ including nuclear recoil corrections and
  the electron anomalous magnetic moment at the level of the
  Breit-Pauli Hamiltonian; $\alpha^{-1} = 137.03599911(46)$, 
  $m_e/m_\alpha = 1.37093355575(61) 10^{-4}$, 
  Ry $c = 3.289841960360(22)\,10^{15}$ Hz. Not indicated is
  the uncertainty due to $\alpha$, which is 0.20 kHz for $\nu_{01}$.
  The last row includes the most recent experimental values.}
\label{table3}
\begin{ruledtabular}
\begin{tabular}{lw{5.4}w{5.8}w{5.0}}
                    &\nu_{01}[{\rm kHz}]  &\nu_{12}[{\rm kHz}] &\mbox{\rm Ref.} \\ \hline
$E_Q$               &           21.73     &          7.42      &                \\
$E_H$               &           -4.21     &          4.05      &                \\
$E_S$               &           11.37(02) &         -1.25(01)  & \mbox{\protect\cite{Pach2}}   \\
$E_L$               &          -29.76(16) &        -12.51(27)  & \mbox{\protect\cite{Pach2}}   \\ \hline
$E^{(7)}$           &           -0.87(16) &         -2.30(27)  &                \\
$E^{(7)}_{\rm log}$ &           82.59     &         -10.09     & \mbox{\protect\cite{log}}     \\
$E^{(6)}$           &        -1557.50(06) &       -6544.32(12) & \mbox{\protect\cite{Drake,yan}, \protect\cite{Pach3,Pach4}}\\
$E^{(4)}$           & 29\,618\,418.79(01) & 2\,297\,717.84     & \mbox{\protect\cite{Drake},\protect\cite{Pach3}}\\ \hline
total               & 29\,616\,943.01(17) & 2\,291\,161.13(30) &                \\
Drake               & 29\,616\,946.42(18) & 2\,291\,154.62(31) & \mbox{\protect\cite{Drake}} \\ 
exp.                & 29\,616\,951.66(70) & 2\,291\,175.59(51) &
\hspace*{1ex}\mbox{\protect\cite{gab,Hessels1,Hessels2,Inguscio1,Inguscio2,shiner}}
\end{tabular}
\end{ruledtabular}
\end{table}

Since all relevant contributions to helium fine structure splitting 
now seem to be known, we are at a position
to present final theoretical predictions, which is done in Table III.
Although we have included all terms up to order $m\,\alpha^7$,
theoretical predictions are in apparent disagreement with
the measurements, as can be seen from the last row of Table \ref{table3}.
Let us analyze possible sources
of this discrepancy. The numerical calculation involves
variational nonrelativistic wave function. The parameter which controls
its accuracy is the nonrelativistic energy. Our wave function,
consisting at maximum of 1500 explicitly correlated exponential functions
reproduces energy with 18 significant digits 
in agreement with the result of Drake in \cite{Drake}. 
Matrix elements with this wave function
are not as accurate as nonrelativistic energy, but they are sufficiently
accurate for leading fine structure operators, and results agree with 
more accurate and independent calculation of Drake  in \cite{Drake}.
For example, $E^{(4)}$ agrees to $0.01$ kHz and $E^{(6)}$ to $0.1$ kHz.
In fact almost all numerical calculations have been performed by us 
and by Drake independently with one exception; we have not obtained
recoil correction to the second order matrix element with Breit operators 
in $E^{(6)}$. More important is 
the complexity of the derivation of $m\,\alpha^7$ operators,
namely $H_i$ and $Q_i$. We purposely derived $H_i$ 
in a way very similar way to the derivation of the D-K operators to avoid accidental mistakes. 
We note that the $Q_i$ operators were obtained  from the one-loop scattering amplitude 
in an almost automatic way, in contrast to the former very lengthy derivation 
of Zhang \cite{tao1,tao2,tao3}, with which we are in disagreement
(see summary of Zhang results in Ref. \cite{Drake}). 
In our previous papers with Sapirstein
\cite{Pach2, Pach4} we pointed out several computational mistakes and inconsistencies
in Zhang's calculations, and therefore we consider the result of Drake,
see Table \ref{table3}, to be incomplete.
While it is possible that we have made a mistake somewhere,
the other probable explanation of the discrepancy with experiments 
is the neglect of higher order terms, namely $m\,\alpha^8$. 
An indication of their importance is the recoil correction
to the second order contribution, obtained by Drake in \cite{Drake}. 
In spite of the small electron-alpha particle mass ratio, 
this correction is very significant; for example $\delta\nu_{01} = -10.81$
kHz. The mass ratio $m_e/m_\alpha \approx 0.00014$ is not much different from 
$\alpha^2 \approx 0.000053 $, and for this reason one can expect that iteration of Breit-Pauli
Hamiltonian in the third order might also be significant.
However, most $m\,\alpha^8$ operators should be negligible,
as $E^{(7)}$ is already at the few kHz level, so that an additional power 
of $\alpha$ will make these operators contribute well below the experimental accuracy.

In summary, we have obtained the complete $\alpha^5$ Ry 
contribution to helium fine structure splitting. Theoretical predictions,
including this result, are in disagreement with measurements 
\cite{gab, Hessels1, Hessels2, Inguscio1, Inguscio2,shiner}. Therefore
the determination of $\alpha$ from helium spectroscopy requires
both checking the calculation of $E^{(7)}$ and the reliable estimation
of the higher order  $E^{(8)}$ contribution, which is a challenging task.
Therefore, at present, helium fine structure splitting is not competitive 
with respect to other determinations of $\alpha$, for example, from the recent 
experiment on the photon recoil \cite{biraben}. 

I am grateful to Andrea Ferroglia and Micha\l\ Czakon for useful advice
on evaluation of d-dimensional Feynman diagrams.

\end{document}